\documentclass[12pt,a4paper,twoside]{paper}
\usepackage{amssymb}
\usepackage{graphicx}
\usepackage{epstopdf}

\newcommand{\be}{\begin{equation}}
\newcommand{\ee}{\end{equation}}

\newcommand{\ba}{\begin{eqnarray}}
\newcommand{\ea}{\end{eqnarray}}

\newcommand{\bea}{\begin{eqnarray*}}
\newcommand{\eea}{\end{eqnarray*}}

\newcommand{\bee}{\begin{enumerate}}
\newcommand{\ene}{\end{enumerate}}

\usepackage{epsfig}
\def\R{\mathbb R}

\ifx\optionkeymacros\undefined\else \fi

\catcode`\Œ=\active\defŒ{{\aa}}       
\catcode`\º=\active\defº{\int}        
\catcode`\=\active\def{\c c}        
\catcode`\¶=\active\def¶{\partial}    
\catcode`\Ä=\active\defÄ{\oint}       
\catcode`\Æ=\active\defÆ{\triangle}   
\catcode`\Â=\active\defÂ{\neg}        
\catcode`\µ=\active\defµ{\mu}         
\catcode`\¿=\active\def¿{{\o}}        
\catcode`\¹=\active\def¹{\pi}         
\catcode`\Ï=\active\defÏ{{\oe}}       
\catcode`\§=\active\def§{{\ss}}       
\catcode`\ =\active\def {\dagger}     
\catcode`\Ã=\active\defÃ{\sqrt}       
\catcode`\·=\active\def·{\Sigma}      
\catcode`\Å=\active\defÅ{\approx}     
\catcode`\½=\active\def½{\Omega}      
\catcode`\£=\active\def£{{\it\$}}     
\catcode`\°=\active\def°{\infty}      
\catcode`\¤=\active\def¤{{\S}}        
\catcode`\¦=\active\def¦{{\P}}        
\catcode`\¥=\active\def¥{\bullet}     
\catcode`\»=\active\def»{\leavevmode\raise.585ex\hbox{\b a}}      
\catcode`\¼=\active\def¼{\leavevmode\raise.6ex\hbox{\b o}}        
\catcode`\­=\active\def­{\not=}       
\catcode`\²=\active\def²{\leq}        
\catcode`\³=\active\def³{\geq}        
\catcode`\Ö=\active\defÖ{\div}        
\catcode`\É=\active\defÉ{{\dots}}     
\catcode`\¾=\active\def¾{{\ae}}       
\catcode`\Ç=\active\defÇ{\ll}         
\catcode`\Ò=\active\defÒ{``}          
\catcode`\Á=\active\defÁ{!`}          
\catcode`\¢=\active\def¢{\rlap/c}     
\catcode`\Ô=\active\defÔ{`}           
\catcode`\Õ=\active\defÕ{'}           

\catcode`\=\active\def{{\AA}}       
\catcode`\'=\active\def'{\c C}        
\catcode`\¯=\active\def¯{{\O}}        
\catcode`\¸=\active\def¸{\Pi}         
\catcode`\Î=\active\defÎ{{\OE}}       
\catcode`\®=\active\def®{{\AE}}       
\catcode`\×=\active\def×{\diamond}    
\catcode`\¡=\active\def¡{\accent'27}  
\catcode`\Ó=\active\defÓ{''}          
\catcode`\±=\active\def±{\pm}         
\catcode`\È=\active\defÈ{\gg}         
\catcode`\À=\active\defÀ{?`}          
\catcode`\Ð=\active\defÐ{--}          
\catcode`\Ñ=\active\defÑ{---}         

\catcode`\Š=\active\defŠ{\"a}        
\catcode`\'=\active\def'{\"e}        
\catcode`\•=\active\def•{\"{\i}}     
\catcode`\š=\active\defš{\"o}        
\catcode`\Ÿ=\active\defŸ{\"u}        
\catcode`\Ø=\active\defØ{\"y}        
\catcode`\€=\active\def€{\"A}        
\catcode`\…=\active\def…{\"O}        
\catcode`\†=\active\def†{\"U}        
\catcode`\‡=\active\def‡{\'a}        
\catcode`\Ž=\active\defŽ{\'e}        
\catcode`\'=\active\def'{\'{\i}}     
\catcode`\—=\active\def—{\'o}        
\catcode`\œ=\active\defœ{\'u}        
\catcode`\ƒ=\active\defƒ{\'E}        
\catcode`\ˆ=\active\defˆ{\`a}        
\catcode`\=\active\def{\`e}        
\catcode`\"=\active\def"{\`{\i}}     
\catcode`\˜=\active\def˜{\`o}        
\catcode`\=\active\def{\`u}        
\catcode`\Ë=\active\defË{\`A}        
\catcode`\‹=\active\def‹{\~a}        
\catcode`\–=\active\def–{\~n}        
\catcode`\›=\active\def›{\~o}        
\catcode`\Ì=\active\defÌ{\~A}        
\catcode`\"=\active\def"{\~N}        
\catcode`\Í=\active\defÍ{\~O}        
\catcode`\‰=\active\def‰{\^a}        
\catcode`\=\active\def{\^e}        
\catcode`\"=\active\def"{\^{\i}}     
\catcode`\™=\active\def™{\^o}        
\catcode`\ž=\active\defž{\^u}        

\let\optionkeymacros\null

\begin{document}

     \vspace*{7mm}

     \vspace*{10mm}

     \begin{center}
 \textsl{\LARGE    Diagnosing the Trouble With Quantum Mechanics }

 \vspace*{10mm}

 { \Huge Jean Bricmont\footnote{IRMP,
Universit\'e catholique de Louvain,
chemin du Cyclotron 2,
1348 Louvain-la-Neuve,
Belgium. E-mail: jean.bricmont@uclouvain.be},
Sheldon Goldstein\footnote{Department of Mathematics, Rutgers University, Hill Center, 110 Frelinghuysen Road, Piscataway,
NJ 08854-8019, USA. E-mail: oldstein@math.rutgers.edu}

\vspace*{10mm}
  { \Large  Dedicated to the memory of Pierre Hohenberg}
 
 }
\end{center}

\begin{abstract}

We discuss an article by Steven Weinberg \cite{W} expressing his discontent with the usual ways to understand quantum mechanics.
We examine the two solutions that he considers and criticizes and propose another one, which he does not discuss, the pilot wave theory or Bohmian mechanics, for which his criticisms do not apply. 

\end{abstract}

\newpage

\vspace*{5mm}

\section{Introduction: Weinberg's Trouble with Quantum Mechanics}\label{sec1}

Pierre Hohenberg \cite{PH, PH1}, along with many other very distinguished physicists,  including Murray Gell-Mann and  Jim  Hartle \cite{GMH}, Robert Griffiths \cite{RG}, Roland Omn\`es \cite{Om}, Gerard t'Hooft \cite{GH}, and J\"urg Fr\"ohlich \cite{JF, JF2}, have expressed their discontent with our present understanding (or lack thereof) of quantum mechanics. We will discuss here the views of one preeminent discontented physicist, Steven Weinberg, who has stated clearly and unambiguously that there is
 something rotten in the kingdom of the ``Copenhagen interpretation" of quantum mechanics.  Though most physicists  often continue to pay lip service to that ``interpretation" in their courses and papers, few of them  are quite sure nowadays what that interpretation really says.
 As Weinberg puts it, ``It is a bad sign that those physicists today who are most comfortable with quantum mechanics do not agree with one another about what it all means." \cite{W}.

Weinberg identifies the basic problem with quantum mechanics: that one applies a different rule of evolution to the wave function of a system when it does not contain an observer or measuring device (one then uses the deterministic Schr\"odinger evolution) from when it does (one then sometimes collapses the wave function in a manner consistent with a probability rule, due to Max Born, for the result).  We agree with Professor Weinberg that this is deeply unsatisfactory, since it seems to ascribe to observers and  measuring devices a fundamental role in physics.

Weinberg is also dissatisfied with what he calls the ``instrumentalist" view of quantum mechanics, which rejects that theory as a description of reality and views it as ``merely an instrument that provides predictions of the probabilities of various outcomes when measurements are made" \cite{W}. Weinberg adds that  ``the trouble with this approach is not only that it gives up on an ancient aim of science: to say what is really going on out there," but that  ``humans are brought into the laws of nature at the most fundamental level" \cite{W}.

The  problem that bothers Weinberg was memorably  expressed  by John Bell some time ago:

\begin{quote}

 It would seem that the theory is exclusively concerned about ``results of
measurement", and has nothing to say
about anything else. What exactly qualifies some physical systems to play
the role of ``measurer"? Was the
wavefunction of the world waiting to jump for thousands of millions of
years until a single-celled living
creature appeared? Or did it have to wait a little longer, for some better
qualified system$\ldots$ with a Ph D?

\begin{flushright} John S. Bell  \cite[p. 34]{Be2}     \end{flushright}

\end{quote}

Does physics really have nothing to say about the world outside of laboratories?
And if that is the case, why build laboratories in the first place?
Theories are not designed merely to account for experiments; rather experiments are designed to serve as rigorous tests of theories.
 Again, Bells says it clearly:

\begin{quote}
But experiment is a tool. The aim remains: to understand the world. To restrict quantum mechanics to be exclusively about piddling laboratory operations is to betray the great enterprise. A serious formulation [of quantum mechanics] will not exclude the big world outside the laboratory.

\begin{flushright} John Bell \cite[p. 34]{Be2}  \end{flushright}
 
\end{quote}

So, Weinberg demands an approach to quantum mechanics in which no  special status is given to the observer in the fundamental laws of physics (and we agree with him about that) and he sees two ways to achieve that goal.

\section{ Weinberg's Two ``Realist" Solutions}\label{sec2}

Weinberg considers two possible solutions to the problems of quantum mechanics:
 the ``many worlds interpretation" (MW) of Hugh Everett  \cite{Eve, DWG} and the ``spontaneous collapse" theories of GianCarlo Ghirardi, Alberto Rimini, and Tullio Weber \cite{GRW, BG}.

In the first approach, the wave function always evolves according to the deterministic Schr\"odinger equation.
But if the wavefunction describes everything that physically exists, that assumption
 has  very strange implications, as Weinberg notes.
 
To understand why,  consider a very simple quantum system, composed of the spin of a single particle, whose initial state is:

\ba
c_1 |1\uparrow \rangle + c_2 |1\downarrow \rangle,
\label{micro}
\ea
where $|1\uparrow \rangle$ denotes the  positive spin state (or ``spin up") in a given direction labelled $1$ and $|1\downarrow \rangle$ denotes the  negative spin state (or ``spin down")  in that direction; $c_1$, $c_2$ are complex numbers with
$|c_1|^2+|c_2|^2=1$.

Consider a measuring device idealized by a pointer that will be up at the end of the measurement if the spin is positive (corresponding to  a wave function $\varphi^\uparrow(z)$) and down if the spin is negative (corresponding to  a wave function $\varphi^\downarrow(z)$). The variable $z$ here is some collective variable indicating the position of the pointer. 

Then it follows simply from the linearity of  the Schr\"odinger evolution that the combined state of the particle and the measuring device, after the measurement, is
 \ba
c_1  \varphi^\uparrow(z)|1\uparrow \rangle + c_2 \varphi^\downarrow (z) |1\downarrow \rangle. \label{macro}
\ea

Since, by further cascading interactions with the measuring device, the rest of the world eventually becomes entangled with the particle in this way, we obtain
  the following situation, described by 
Weinberg: ``the wave function becomes a superposition of two terms, in one of which the electron spin is positive and everyone in the world who looks into it thinks it is positive, and in the other the spin is negative and everyone thinks it is negative" \cite{W}. 

The ``many world interpretation" simply postulates that, in this situation, the world  has split into two ``worlds," each one corresponding to  a term in (\ref{macro}). But of course that happens ``every time a macroscopic body becomes tied in with a choice of quantum states" \cite{W}, which is bizarre to say the least.

Weinberg also observes that for MW it is hard to understand how Born's rule arises, according to which,  upon repetition of the spin measurement of particles with the wave function (\ref{micro}), one should see the spin being positive a fraction $|c_1|^2$ of the time and negative a fraction $|c_2|^2$ of the time.
As Weinberg notes, there have been many attempts to do so, since the 1957 paper of Everett \cite{Eve}, but ``without final success" \cite{W}.

In the spontaneous collapse theories \cite{GRW, BG}, the wave function undergoes, at random times and places, a spontaneous collapse whose main effect is to suppress macroscopic superpositions like (\ref{macro}), which collapse quickly onto one of the terms, thus avoiding the proliferation of ``worlds" of the MW approach.

The difference between the collapse in ordinary quantum mechanics and  in a  spontaneous collapse theory,  is that, for the latter, the collapse is incorporated in the basic equations of motion of the theory (thus, Schr\"odinger's equation is modified) and is not restricted to what happens when measuring devices are involved.

However, since collapses occur much more frequently than in the standard account,
 the predictions of spontaneous collapse theories do not coincide with the usual quantum ones.  Many experiments are, in fact, being carried out in order to decide between spontaneous collapse theories and standard quantum mechanics.   So far, there is no indication that standard quantum mechanics fails predictively, and its spectacular successes so far suggest that, if its predictions are indeed violated in some situations, this will not be easy to demonstrate. 

In addition, the parameters of the spontaneous collapse theory (the frequency and the nature of the collapses) are adjusted in an ad hoc fashion in order that its predictions not deviate from the standard ones for presently feasible experiments, which is not an appealing move to say the least.

\section{A Third Way}\label{sec3}

Weinberg is not satisfied with the instrumentalist approach, where the wave function is not regarded as something to be taken seriously as real or objective, but merely as a convenient tool for describing the behavior of measuring devices and the like; but he is not quite happy either with the two approaches (the many-worlds and spontaneous collapse theories), where  the wave function not only represents something real and objective but is also exhaustive, providing a complete description of the physical state of affairs\footnote{See however Section \ref{sec6} for a serious qualification of that idea.}.

In other words, the alternatives for Weinberg (both being unsatisfactory for him) are
that the wave function is either nothing or everything. However, Weinberg does not mention a third possibility,  that the wave function is something
but not everything. This is the case, for example, with Pierre Hohenberg's version of quantum mechanics \cite{PH, PH1}, which however, has not been sufficiently developed to properly assess its viability. It is also the case with the de Broglie-Bohm theory
or pilot-wave theory or Bohmian mechanics.

This theory, which we consider to be, by far, the simplest version
of quantum mechanics, does not require any modification of the predictions of ordinary
quantum mechanics, nor a bizarre multiplication of parallel universes. It was proposed
by Louis de Broglie in 1927 and rediscovered and developed by David Bohm in 1952. For
several decades its main proponent was John Stewart Bell \cite{Be}, the physicist who did more
than any other to establish the existence of quantum non-locality.

In Bohmian mechanics, a system of particles is described by the actual positions of actual 
particles in addition to its wave function or quantum state: particles that have positions at all times and hence 
 trajectories and also velocities. As we shall explain below, their time evolution is guided in a natural way by the wave function, which functions as what is often called a pilot wave (see Subsection \ref{3sec1}). This should be contrasted with the role of the wave function in the instrumentalist approach: to predict the behavior of (clearly non-fundamental) measuring devices. Thus the wave function in Bohmian mechanics is somewhat similar to the forces or the electromagnetic waves guiding the particles in classical physics. 

The wave function of a closed system in Bohmian mechanics, even a system containing observers and measuring devices, always follows Schr\"odinger's equation and never collapses. Thus, observations are no longer a {\it deus ex machina} in that theory. When one analyzes in Bohmian mechanics what is called a `measurement' in ordinary quantum mechanics, one finds that the behavior of the particles conforms precisely to the quantum mechanical predictions (see Sections \ref{sec4}, \ref{sec5}). Such an analysis of quantum measurements also explains why the fact that particles have both positions and velocities at all times  in Bohmian mechanics does not contradict the Heisenberg uncertainty principle. 

Although Bohmian mechanics is perfectly deterministic, one can recover the statistical predictions of ordinary quantum mechanics (the Born rule mentioned by Weinberg) by making natural assumptions on the initial conditions of physical systems (something which has become familiar among physicists with the development of modern `chaotic' dynamical systems theory), see Subsection \ref{3sec2}. 

In what follows, we will only sketch how Bohmian mechanics  works; for more detailed but still elementary introductions to that theory, see \cite{B2, Tu} and for more advanced ones, see \cite{Bo, BH, B, DGZ, DT,  DGZ1, Go, No}
\footnote{There are also pedagogical videos made by students in Munich, available
at: https://cast.itunes.uni-muenchen.de/vod/playlists/URqb5J7RBr.html.}.

 \subsection{The Equations of Bohmian Mechanics}\label{3sec1}

In Bohmian mechanics, the complete state of a closed physical system composed of $N$ particles is a pair 
$(|$quantum state$>$, $\bf X)$, where  $|$quantum state$>$ is the usual quantum state (the tensor product of wave functions with some possible internal states), and 
${\bf X}= (X_1,\ldots,X_N)$ represents the  positions of the particles that always exist, independently of whether one looks at them or one  measures them (each $X_i\in \R^3$). 

The time evolution of the complete physical state is composed of two laws:

\par\noindent
1.  The usual evolution of the    $|$quantum state$>$, for all times,
{\it whether one measures something or not}. If the state is only a wave function, $|$quantum state$>$= $\Psi(x_1, \dots, x_N)$, with each
$x_i=(x_{i1}, x_{i2}, x_{i3})\in \R^3$, it obeys  Schr\"odinger's equation:

\ba
i  \partial_t \Psi =  {\cal H} \Psi
 \label{S}
\ea

where

\ba
 {\cal H}=-\frac{1}{2} \Delta +V
 \ea
is the quantum  Hamiltonian with, to simplify, $\hbar=1$,  all the masses equal to $1$ and with
$\Delta = \sum_i^N \Delta_i$,
 where 
 $\Delta_i
=\frac{\partial^2}{\partial x_{i1}^2} + \frac{\partial^2}{\partial x_{i2}^2}   + \frac{\partial^2}{\partial x_{i3}^2}$.

\bigskip 

\par\noindent
2.
The particle positions  ${\bf X}
={\bf X} 
(t)$ evolve in time according to a guiding equation determined by the quantum state: their velocity is a function of the wave function. If one writes\footnote{We use lower case letters for the generic arguments of the wave function and upper case ones for the actual positions of the particles.}: 
$$\Psi (x_1, \dots, x_N)=R (x_1, \dots, x_N)e^{iS (x_1, \dots, x_N)},
$$ then:

 \ba
 \frac{ d X_k (t)}{dt}=   \displaystyle  \nabla_k S (X_1(t),\ldots,X_N(t)),
 \label{P}
 \ea
where $ \nabla_k$ is the gradient with respect the coordinates of the  $k$-th particle.
More generally, for quantum states that are multi-component wave functions (corresponding, for example, to particles with ``spin"):
\ba
\frac{ d X_k(t)}{dt} = v^\Psi_k ({\bf X}(t))= \displaystyle   \quad \frac{\textrm{Im} (\Psi^* \cdot\nabla_k \Psi)}{\Psi^\ast \cdot \Psi} (X_1(t),\ldots,X_N (t)).
 \label{P1}
 \ea
where $\cdot$ stands for the scalar product between the components of the quantum state.

The origin of this equation is not mysterious; it is of the form:
\ba
\frac{ d {\bf X}}{dt} = \frac{\bf J}{\rho}
 \label{C}
\ea
where ${\bf J}= \textrm{Im} (\Psi^* \cdot{\bf \nabla } \Psi)$ is the current associated with the ``conservation of probability" in quantum mechanics
and $\rho= |\Psi|^2$ is the quantum probability density for the configuration of positions; ${\bf J}$ and  $\rho$ are related by

 \ba
 \frac{\partial \rho}{\partial t} + \nabla \cdot {\bf J}=0,
  \label{C1}
 \ea
which follows easily from Schr\"odinger's equation (\ref{S}).

 \begin{figure}[t]
 \includegraphics[keepaspectratio,height=10cm]{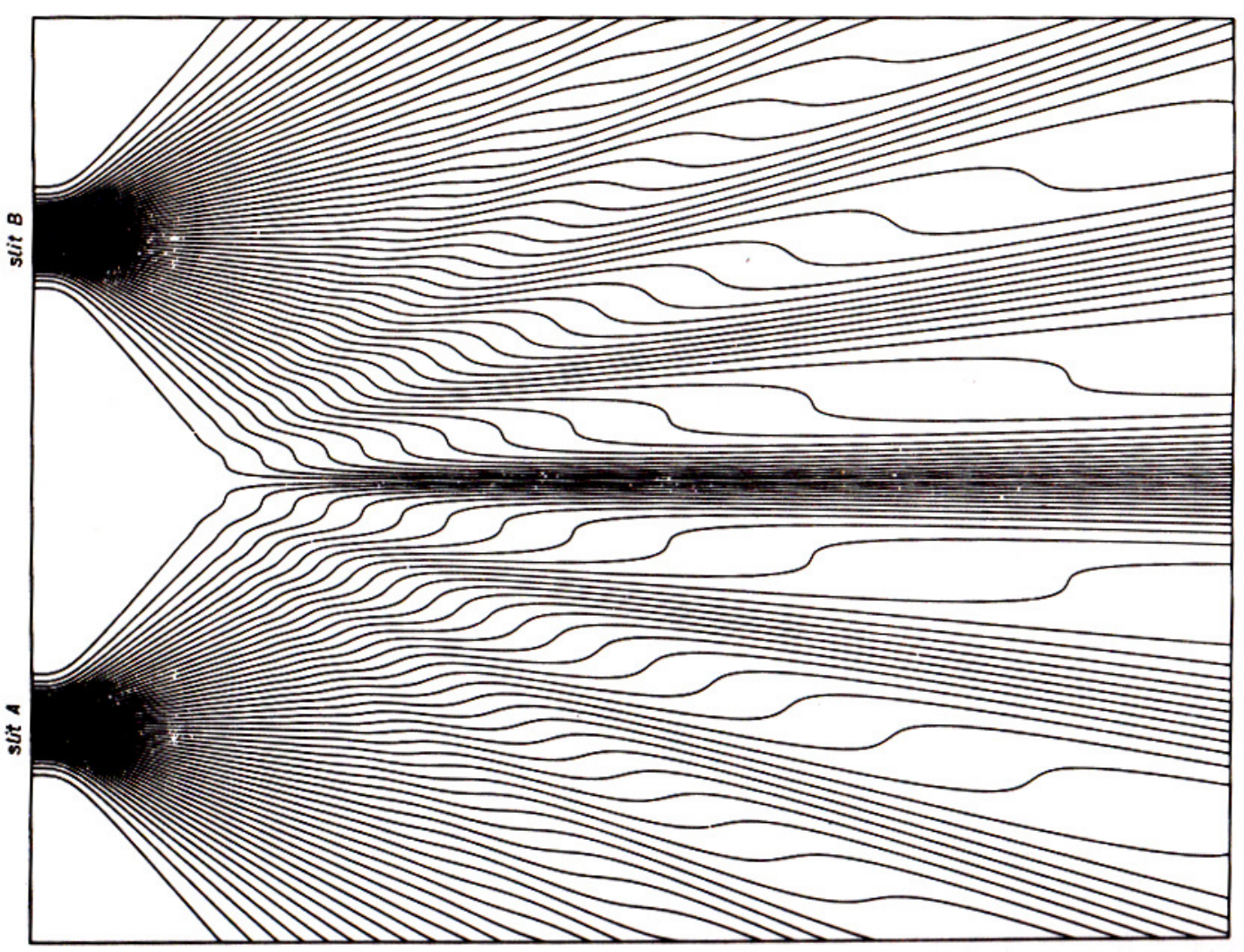}
\caption[]{Trajectories of  the Bohmian dynamics in the two-slit experiment. Each line corresponds to a possible trajectory of a particle, corresponding an initial position just behind one of the slits. Reproduced with the kind permission of {\it Societ\`a Italiana di Fisica\/} and the authors \cite{PDH}}
\label{3fig2}
\end{figure}

This dynamics is best illustrated by the two-slit experiment:  Figure \ref{3fig2} shows a numerical solution of the Bohmian  dynamics for  that experiment.
Note that the motion 
 {\it in vacuum} behind the slits is highly  {\it non classical} ! Newton's first law (rectilinear motion in the absence of  forces) is not satisfied.

 Note also that, if one assumes, as in Figure \ref{3fig2}, that there is a symmetry in the wave function  $\Psi$ between the top and the bottom of the figure, then one can determine a posteriori through which slit the particle went! 
 Indeed, because of the symmetry of the wave function  $\Psi$, its gradient  is tangent to the line in the middle of  Figure \ref{3fig2}; thus, because of  (\ref{P}), the velocity of the particle is also tangent to that line and the particles cannot go through it\footnote{
It is interesting to compare this numerical solution to results published in  
 {\it Science} in June 2011 \cite{Koc}: one finds that the profile of
trajectories of photons obtained through a series of so-called ``weak measurements" is qualitatively similar to that of 
  Figure \ref{3fig2}.}.

\subsection{The Statistical Assumptions in Bohmian Mechanics}\label{3sec2}

In order to understand why Bohmian mechanics reproduces the usual quantum predictions, one must use a fundamental consequence of that dynamics, 
{\it equivariance}:  If the probability density $\rho_{t_0}({\bf x})$ for the initial configuration $ {\bf X}_{t_0}$
 is given by $\rho_{t_0}({\bf x}) = |\Psi ({\bf x}, t_0)|^2$, then the probability density for the configuration ${\bf X}_t$ at any time t is given by 
\ba
 \rho_t ({\bf x})= |\Psi ({\bf x}, t)|^2,
  \label{QE}
\ea
where $\Psi  ({\bf x}, t)$ is a solution to the  Schr\"odinger's equation (\ref{S}).

\begin{figure}[t]
\centering
\includegraphics[keepaspectratio,height=8cm]{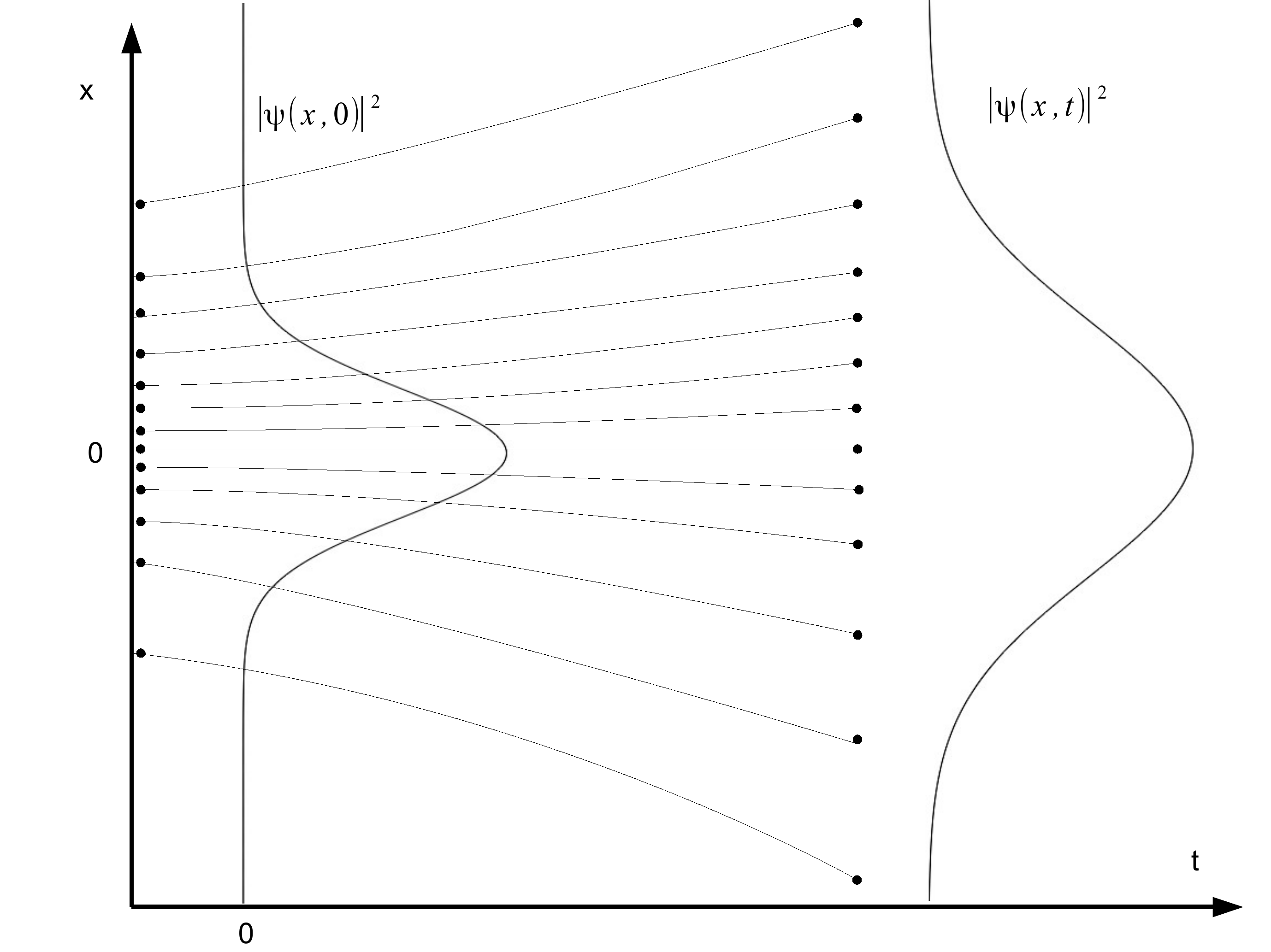}
\caption[]{Illustration of the equivariance  of Bohmian mechanics}\label{3fig3}
\end{figure}

This follows easily from equations (\ref{C}, \ref{C1}) and is illustrated  by 
 Figure \ref{3fig3}, where  each curve describes a trajectory (for simplicity, we consider in Figure \ref{3fig3} a one-dimensional system). The figure depicts a random distribution of initial positions whose density is approximately given by $ |\Psi  (x, 0)|^2$ (where the variable $x \in \R$ is on the vertical axis)  and one sees that at a later time  $t$ this density will approximately be given  by $|\Psi  (x, t)|^2$.

Because of equivariance, the quantum predictions for the results of   measurements of any quantum observable  are
obtained  if one assumes that the initial density satisfies $\rho_{t_0}  ({\bf x}) = |\Psi  ({\bf x}, t_0)|^2$. The assertion
that configurational probabilities at any time $t_0$ are given by this ``Born rule" is called
the {\it quantum equilibrium hypothesis}. The justification of the quantum equilibrium hypothesis -- and, indeed, a clear  statement of what it actually means -- is a long story, too long to be discussed here (see \cite{DGZ})\footnote{Equation (\ref{QE}) is  useful primarily when it is applied to  subsystems of a larger system, for example the universe, that has its own wave function. In that case, one can associate to the subsystem an effective wave function $\Psi$ and the empirical distribution $\rho$ of particle configurations in appropriate ensembles of subsystems,  each having effective wave function $\Psi$,  is given by (\ref{QE}).}.

Thus, in Bohmian mechanics, the quantum state $\Psi$ has a double status:

\par\noindent
--  It generates, through equations (\ref{P}) or (\ref{P1}), the motion of particles.

\par\noindent
--   It also governs the statistical distribution of  configurations of particles, via $|\Psi|^2$.

One may compare the quantum state $\Psi$ with the Hamiltonian $H$ in classical physics. The latter generates the motion of particles, through Hamilton's equation, and  also gives the statistical distribution at equilibrium. Formally, the analogy is as follows: 
 $ {\cal H} \sim -\log \Psi$ and  $ |\Psi|^2\sim \exp (-\beta  {\cal H})$, with $\beta=2$.

The fact that the only variables introduced in  Bohmian mechanics, beyond the wave function,  are the positions of particles, which change in time, is sometimes regarded as an argument against that theory: Why focus so rigidly on positions and not also on other observables?  But a moment's reflection shows  that the situation is completely similar to that in classical mechanics: all quantities of physical interest, like velocities, energies and angular momenta, are functions of the positions of the particles and their motions. 

And, in Bohmian mechanics, the values of all ``quantum observables"  other than positions can be derived from an analysis of the particles' trajectories, as we shall illustrate in two concrete examples in the following sections, involving  spin and  momentum.

 \section{The ``measurement" of spin in the  Bohmian Mechanics}\label{sec4}

In Bohmian mechanics, spin is not real. What we mean by this is that, unlike position, corresponding to which the wave function in Bohmian mechanics is supplemented with determinate values $X_i$, no such supplementation is made to deal with spin. Instead the wave function gets additional components. Nonetheless, in Bohmian mechanics, as in standard quantum mechanics, there are ``spin measurements."

\begin{figure}[t]
\centering
\includegraphics[width=.9\textwidth]{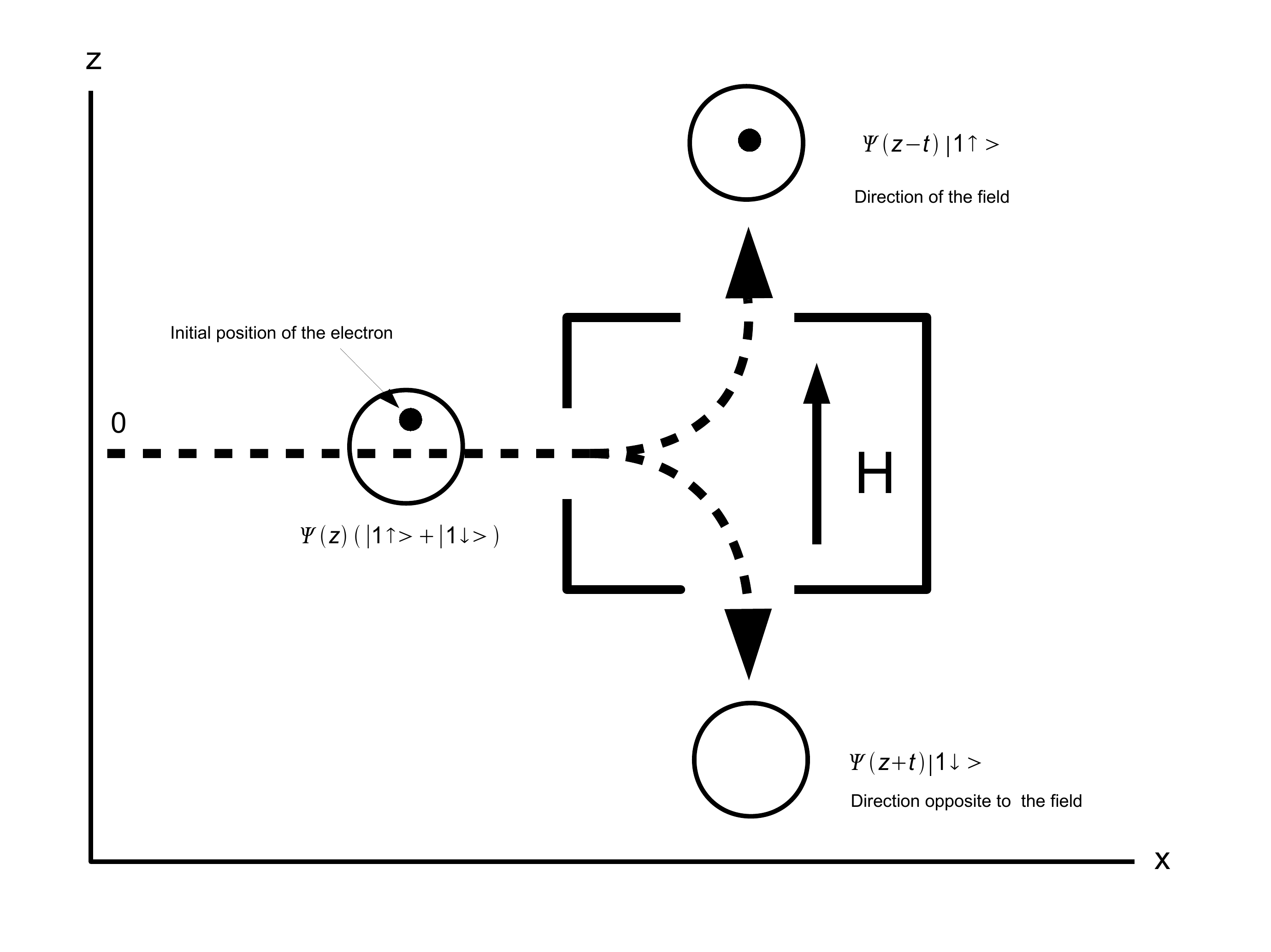}
\caption[]{An idealized spin measurement}\label{3fig4}
\end{figure}

 In order to understand spin measurements in Bohmian mechanics, consider  a quantum state $\Phi$ which is a superposition of
$|1 \uparrow>$ and $|1 \downarrow>$, that we describe in an idealized form\footnote{This Section is based on Chapter 7 of David Albert's book ``Quantum Mechanics and Experience" \cite{Al}.}:  
\ba
\Phi= \Psi(z) (|1 \uparrow> +|1 \downarrow>)
\ea 
$z$ being the vertical direction  (see Figure \ref{3fig4}). We will assume that the spatial part of the state, namely the wave function  $\Psi(z)$, is symmetrical: $\Psi(z)= \Psi(-z)$. That implies that the line  $z=0$ is a nodal line, through which the gradient of  $\Psi(z)$ is zero and that the particles cannot cross, by (\ref{P}) (the situation is similar to the one of Figure \ref{3fig2}). $\Psi$ is also a function of the horizontal variable $x$ (the particle  moves rightwards in the $x$ direction), but we suppress that variable.

In Figure \ref{3fig4}, $H$ denotes an inhomogeneous magnetic field; the disks represent (in a very idealized way) the support of the spatial part of the wave function. The part  $|1 \uparrow>$ of the quantum state always goes in the direction of the field (which gives rise to the state $\Psi(z-t) |1 \uparrow>$)
and the part $|1 \downarrow>$ always goes in the direction opposite to the field  ($\Psi(z+t) |1 \downarrow>$).
  
 But the particle, if it starts initially in the part of the wave function above the nodal line $z=0$ (produced by the symmetry of the wave function), will always go up, because it cannot cross that nodal line\footnote{Since the particles here have spin, the guiding equation  is (\ref{P1}), not (\ref{P}), but this does not affect our qualitative discussion.}. We then say that the spin has been measured to be up.

\begin{figure}[t]
\centering
\includegraphics[width=.9\textwidth]{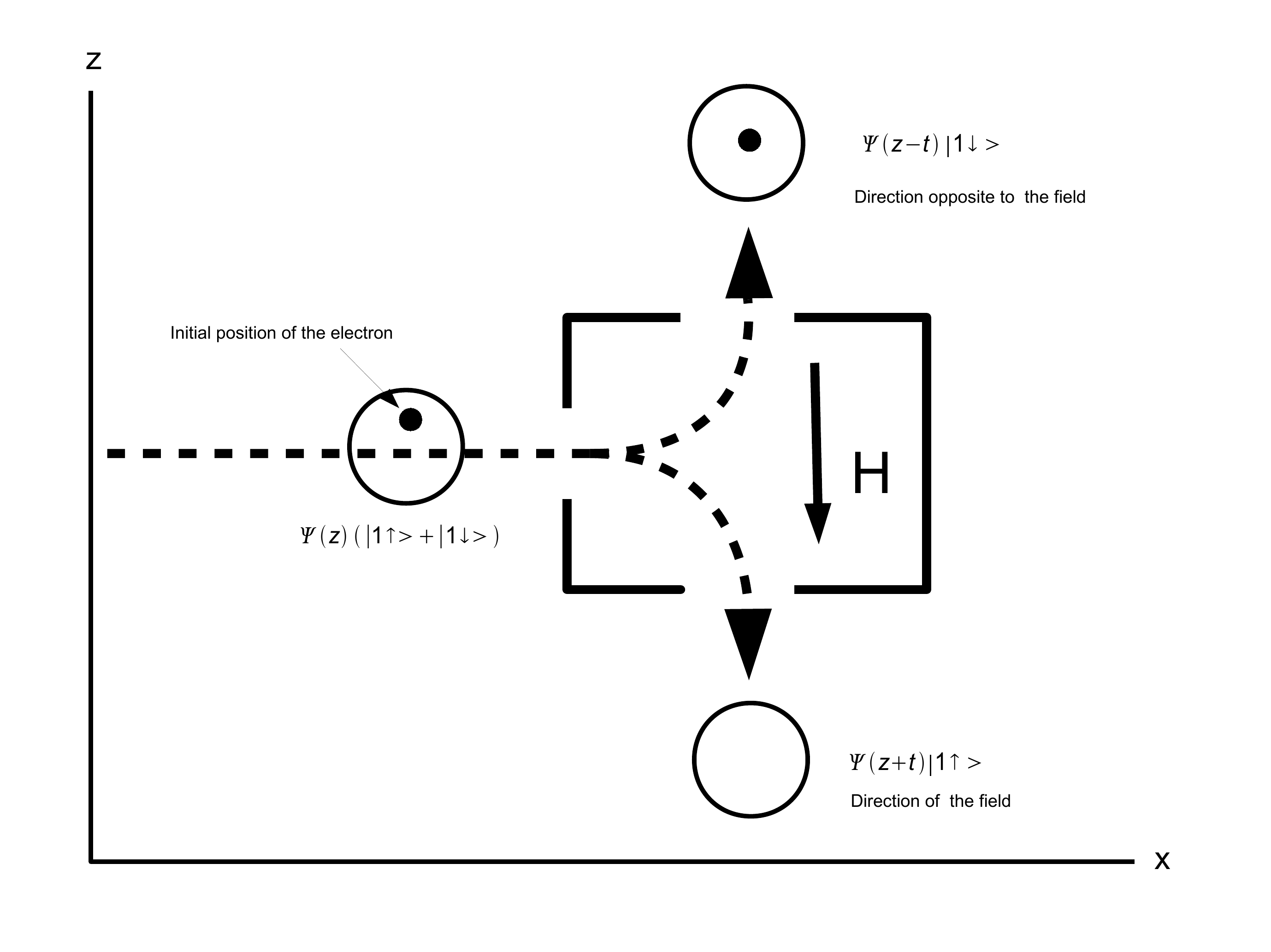}
\caption[]{An idealized spin measurement with the direction of the field reversed with respect to the one of Figure~\ref{3fig4}}\label{3fig5}
\end{figure}

To underline the fact that in Bohmian mechanics there is no actual spin quantity that has been measured in this way, consider reversing the direction  of the field, as in Figure \ref{3fig5}. Since it cannot cross the nodal line, the
 particle, with exactly the same initial position as in  Figure \ref{3fig4}, will again  go upwards. But then, what was  ``positive spin" becomes ``negative spin" (i.e. going in the direction opposite to the one of the field), although one ``measures"  the spin in the  {\it same} direction
in both set-ups, and with the {\it same initial conditions} for the particle (both its wave function and position), with only 
  arrangements of the apparatus differing. 
  
Note that the measuring device is not ``passive" (it does not record  any intrinsic property of the particle  pre-existing to the measurement) but ``active."
 This justifies Bohr's intuition about
 
  \begin{quote}

[\dots ] the {\it impossibility of any sharp distinction between the behavior of atomic objects and the interaction with the measuring instruments which serve to define the conditions under which the phenomena appear}.

\begin{flushright}  Niels Bohr \cite[p.~210]{Bohr},  (italics in the original) \end{flushright}

  \end{quote}
  
But in Bohmian mechanics, this ``intuition" is a consequence of the theory.

Using equivariance, one can show that the statistical predictions for the results of ``spin measurements" in Bohmian mechanics do not depend on the  choice (up or down) of the orientation of the field and coincide with the usual ones.

Note also that both parts of the wave function in  Figures \ref{3fig4} and \ref{3fig5} continue to evolve according to the usual equations. But the particle is guided only by the part of the wave function in the support of which it is located (which follows from eqs. (\ref{P}, \ref{P1})). This means that one can, in practice and in certain cases, reduce the wave function and only keep the part  in the support of which the particle is situated. However, in Figures  \ref{3fig4} and \ref{3fig5}, it could happen that  those two parts of the wave function  overlap later and therefore one cannot forget the part in the support of which the particle does not find itself.
  
 But one can show that, when the particle interacts with a macroscopic apparatus  and  one obtains a state like (\ref{macro}), then the overlapping of the wave functions in the two terms of (\ref{macro}) is in practice impossible,
and one may, again in practice, keep only the part of the wave function in the support of which the particle is if one wants to analyze its future behavior.
In that sense, one   ``reduces" the quantum state but this operation is ``effective," somewhat like the irreversibility of macroscopic laws in classical physics\footnote{In fact, one can introduce a notion of wave function for a subsystem of a closed system (i.e. in principle of the Universe) that coincides with the  wave function used in quantum mechanics and that does collapse when collapses occur according to the standard approach, see \cite{Go5} for a detailed discussion. The wave function that never collapses is the one of the closed system.}.
 
\section{The ``measurement" of momentum in Bohmian Mechanics}\label{sec5}

What about  ``momentum measurements"? In Bohmian mechanics, particles have a velocity at all times. So, unlike spin, particles always have what we would be inclined to call a momentum (mass $\times$ velocity) in Bohmian mechanics. So one might ask, what sort of probabilities does Bohmian mechanics supply for that: will they agree with the quantum mechanical probabilities for momentum? The answer is no!\footnote{There exists also  a no hidden variables theorem, due to Robert Clifton \cite{RC}, preventing us from assigning both a position and a velocity to two particles on a line, in such a way that the statistical
distributions of these quantities and of certain functions of them coincide with the usual quantum predictions. See \cite[p. 43]{B}  for a discussion of that theorem.}

Besides, one may ask: isn't having both a position and a velocity at the same time contradicted by   Heisenberg's inequalities?

To understand what is going on, we should analyze ``momentum measurements," or what are called momentum measurements in standard quantum mechanics.
Consider a simple example, namely a particle in one space dimension with initial wave function: $ \Psi (x, 0)=\pi^{-1/4}\exp(- x^2/2)$. Since this function is real, its phase  $S=0$  and the particle is  at rest (by equation (\ref{P}): $ \frac{d X(t)}{dt} =  \frac{\partial S(X(t),t)}{\partial x}$).
Nevertheless, the measurement of momentum  $p$ must have, according to the usual quantum predictions,  a probability distribution whose density is given by the square of the  Fourier transform of $\Psi (x, 0)$, i.e. by
 $|\hat \Psi (p)|^2=\pi^{-1/2}\exp(- p^2)$.
 
 Isn't there a contradiction here?
 
In order to answer that question, one must focus on the quantum mechanical {\it measurement}
 of momentum. One way to do this is to let the particle evolve and to detect its asymptotic position  $X(t)$ as
   $t\to \infty$. Then, one sets  $p= \displaystyle \lim_{t\to \infty} \frac{X(t)}{t}$
 (putting the mass $m=1$).
 
Consider the free evolution of the initial wave function at $t_0=0$, $ \Psi (x, 0)=\pi^{-1/4}\exp(- x^2/2)$. The solution of  Schr\"odinger's equation ((\ref{S}), with $V=0$) with that initial condition is:

\begin{equation}
\Psi (x,t) = \frac{1}{(1+it)^{1/2}} \frac{1}{\pi^{1/4}}\exp\left[- \frac{x^2}{2 (1+it)}\right],
\label{Ga1}
\end{equation}
and thus
\begin{equation}
| \Psi (x,t)|^2=\frac{1}{\sqrt{\pi \big[1+t^2\big]} }\exp\left[- \frac{x^2}{ 1+t^2}\right].
\label{Ga}
\end{equation}

If one writes $\Psi (x,t)= R (x,t) \exp \big[iS (x,t)\big]$, one gets  (up to a constant in $x$): 
\[
S(x,t)= \frac{t x^2}{2 (1+t^2)}, 
\]
and the guiding equation (\ref{P}) becomes:
\begin{equation}
 \frac{d}{dt} X(t)=  \frac{t X(t)}{1+t^2},
\label{Ga2}
\end{equation}
whose solution is:
\begin{equation}
X(t)=X(0) \sqrt{1+t^2}.
\label{Ga3}
\end{equation}
This gives the explicit dependence of the position of the  particle as a function of time.
If the particle is initially at   $X(0)=0$, it does not move; otherwise, it moves asymptotically, when  $t\to \infty$, as $X(t)\sim X(0)t$.
Thus, $p = \lim_{t\to \infty} X(t)/t = X(0)$.

Now, assume that we start with the quantum equilibrium distribution: 
$$
\rho_0 (x)= |\Psi (x, 0)|^2=\pi^{-1/2}\exp(- x^2).
$$
This is the distribution of $X(0)$.
Thus, the distribution of  $p = \lim_{t\to \infty} X(t)/t = X(0)$ will be
$\pi^{-1/2} \exp (-p^2) = |\hat \Psi(p,0)|^2$. This is the quantum prediction! But the detection procedure (measurement of $ X(t)$ for large $t$) does {\it not }  measure the initial velocity (which is zero for all the particles).

Thus, although the particles do have, at all times, 
 {\it a position and a velocity},  there is no contradiction between
Bohmian mechanics and the quantum predictions and, in particular, with Heisenberg's uncertainty principle. The latter is simply a relation between variances of results of measurements. It implies nothing whatsoever about what exists or does not exist outside of measurements, since those relations  are simply mathematical consequences of the quantum formalism which, strictly speaking, says nothing about the world outside of measurements.

Quantum ``measurements" of most quantum observables  are in reality not genuine measurements but merely {\it interactions} between a microscopic physical system and a macroscopic measuring device. This is as true in standard quantum mechanics as  in Bohmian mechanics. But for Bohmian mechanics there are genuine position measurements. To use a fashionable expression, one might
say for both Bohmian mechanics and standard quantum mechanics,  values of most observables are {\it emergent}. But it is only in Bohmian mechanics that one can understand how that emergence comes about.

 \section{The Need for an ``Ontology"}\label{sec6}

Let us now come back to the two non-instrumentalist options that Weinberg considers. We said that they both assume that ``the wave function is everything." That is the way they are usually thought of by their defenders.

But that cannot be right! Indeed, let us 
consider, in (\ref{macro}), one of the two wave functions associated to the states of pointers: $ \varphi^\uparrow(z)$ or $ \varphi^\downarrow(z)$.

These are not pointers! A pointer (or any other macroscopic object) is something located in three dimensional space and that changes with time. On the other hand, a wave function is  a vector in  a Hilbert space or, more concretely, a function  defined on a high dimensional space (or a vector of such functions), an element of $L^2(\R^{3N})$, where $N$ equals the number of particles that the pointer is composed of (here we introduced in $ \varphi^\uparrow(z)$ and $ \varphi^\downarrow(z)$ only the variable $z$, which is related to the position of the pointer, but, strictly speaking, those wave functions depend on all the degrees of freedom associated to  the particles in the pointer). Such a function  simply does not assign values to points in the ordinary space  $\R^3$!

 In both the many worlds interpretation and the spontaneous collapse theories,  one implicitly identifies wave functions, at least those of macroscopic objects like pointers, with three dimensional objects, but, for the theory to be well-defined,  this identification should be made explicit. 
 
 So, what those  theories lack is an explicit ontology of objects in space-time or, using a term invented by John Bell, ``local beables." The word ontology here, which may scare people as being too philosophical, simply refers to what exists, or, to be more precise, to what is postulated to exist by a physical theory. It could include atoms, elementary particles, stars, or fields. The word ``beable" was invented in order to contrast it with the word ``observable" that is so central to the standard quantum mechanical terminology. ``Beable" refers to what exists, independently of whether we look at it or not. Finally beables are ``local" if they are located in space, like pointers\footnote{For a further discussion of the need for local beables in theories that modify Schr\"odinger's equation in order to produce spontaneous collapses, see \cite{ASTZ}. For a similar discussion in the ``many-worlds" theory of Everett, see \cite{ASTZ1}.}.
 Even the ``instrumentalist" version of  quantum mechanics postulates  the existence of some local beables (without using that expression of course), namely the measuring devices.

So, a major problem of the instrumentalist and the two non-instrumentalist approaches of Weinberg is that they lack a clear ontology. In all these approaches, one implicitly assumes that wave functions like $ \varphi^\uparrow(z)$ or $ \varphi^\downarrow(z)$ corresponds to pointers, but that means that there exists something other than those wave functions. It is a great merit of Bohmian mechanics that it postulates an explicit local ontology and, moreover, one that is not restricted to an ill-defined ``macroscopic scale."

Local beables are an instance of what are often called ``hidden variables," namely variables, in addition to the wave function, that describe the state of a system. 
There exist several ``no hidden variables theorems" that suggest that the introduction of such variables is difficult if not impossible, see \cite{Be1, KS, Me, Peres1, Per}. For example, one cannot assume  that particles have a position and a momentum whose statistical distributions coincide with the quantum mechanical ones \cite{RC}.

One of the main virtues of Bohmian mechanics is that it does introduce ``hidden variables," namely the positions of the particles, but only those, and, by doing so, it avoids being a priori refuted by the no hidden variables theorems.

 Bohmian mechanics shows that measurements of quantum observables other than positions are typically merely interactions with a measuring device whose statistical results coincide with the quantum predictions. They do not  reveal a pre-existing property of the particle, either because that property does not exist (as is the case for the spin, see Section \ref{sec4})
or because, although it does exist, what quantum mechanics calls a  measurement of that property does not reveal its value (as is the case for the momentum, see Section \ref{sec5}). Moreover, with Bohmian  mechanics, the unacceptable instrumentalist understanding of the wave function of the Copenhagen interpretation, as an instrument coordinating results of measurements, is transformed into an acceptable one, as an instrument choreographing, on the fundamental level, the behavior of particles.  

Finally, we note that with Bohmian mechanics most, if not all, of the mysteries of quantum mechanics are eliminated. This is sometimes raised as a criticism of Bohmian mechanics. For example,  Richard Friedberg and Pierre Hohenberg \cite[pp. 319--320]{PH1}  write:

\begin{quote}
The primary motivations for both de
Broglie and Bohm were first to replace the operationalism of Copenhagen with an
ontology which would not depend on ill-defined concepts for its definition, a point of
view with which we clearly sympathise. A second motivation was of course to solve
the measurement problem. In our view, however, the theory -- however ingenious --
is profoundly misguided, since its classical ontology misses the essential physical
elements of QM, which derive from Hilbert space and lend quantum processes and
quantum information their unique and `miraculous' features. It is thus not surprising
to us that the theory has attracted relatively little interest in the physics community.
\end{quote}

We obviously agree with the beginning of this quote.
However we fail to understand how Bohmian mechanics ``misses the essential physical
elements of QM," since there is no quantum phenomenon that is not accounted for and explained by that theory.
And we consider the elimination of the miraculous  one of its  great virtues.

\vspace*{5mm}

{\bf Acknowledgment.} We thank Tim Maudlin for very interesting discussions on the subject of this paper.

\end{document}